\documentclass[runningheads]{llncs}

\usepackage[T1]{fontenc}

\usepackage{graphicx}
\usepackage{booktabs}
\usepackage{multirow}
\usepackage{listings}
\usepackage{xcolor}
\usepackage{amsmath,amssymb}
\usepackage{stmaryrd}
\usepackage{enumitem}
\usepackage{tikz}
\usepackage{booktabs}
\usepackage{tabularx}
\usepackage{array}
\usepackage{comment}
\usepackage{xspace}

\usetikzlibrary{
  positioning,
  arrows.meta,
  fit,
  calc,
  backgrounds,
  shapes.geometric
}

\usepackage{hyperref}

\newcommand{\sys}{\textsc{C-Trace}\xspace}

\setlist{nosep,topsep=2pt,partopsep=0pt}

\lstdefinelanguage{rego}{
  keywords={package, import, default, allow, deny, not, in, with, as, if, else, some, every},
  sensitive=true,
  comment=[l]{\#},
  morestring=[b]",
}

\begin{document}

\title{Runtime Compliance Verification for AI Agents}

\titlerunning{Runtime Compliance Verification for AI Agents}

% --- Double-blind: anonymized for submission. Restore for camera-ready. ---

\author{
Nafiseh Kahani\inst{1}\orcidID{0000-0002-9322-0699}
\and
Masoud Barati\inst{1}\orcidID{0000-0001-7829-2240} \and Diana Addae \inst{1}\orcidID{0009-0009-8833-0488}}

\authorrunning{N. Kahani et al.}
\institute{
Carleton University, Canada\\
\email{\{NafisehKahani,MasoudBarati\}@cunet.carleton.ca, DianaAddae@cmail.carleton.ca}}

\maketitle

\begin{abstract}
AI agents now handle personal data through tool use, function calls, and multi-turn dialogue, which can create obligations under the General Data Protection Regulation (GDPR). Current testing practices mainly rely on offline red-teaming or static prompt review, but they do not guarantee at run time that an agent's behavior follows regulatory rules. We propose \sys (Compliance Trace-based Runtime Agent Conformance Enforcement), a verification framework that
(i)~expresses a subset of GDPR (consent, purpose limitation, data minimization, and the right to erasure) as formal policy predicates over agent execution traces, (ii)~uses a runtime monitr that intercepts every tool invocation and model output and rejects non-compliant actions, and (iii)~tests the agent with attack dialogues, including DSPy-generated prompts and verbatim prompts from red-teaming corpora, that try to induce violations. We evaluate the framework on four case studies reframed to GDPR. Under 10\% per-category extractor noise (drop-out and over-typing), the monitor keeps attack-success rate at $\le 12\%$ (below the baselines we compare against) and false positives at $\le 16\%$, and reaches $0\%$ ASR under perfect extraction.

%Model alignment and Presidio, a production Personally Identifiable Information (PII) classifier, are not enough for these stateful principles because they do not track earlier turns; an independent GDPR-article judge largely agrees with the oracle labels (88\% overall, with consent lower at 71\%). %A sensitivity analysis locates the monitor's real failure mode in the natural-language category extractor rather than in the temporal logic.

\vspace{0.3em}
\end{abstract}

%==================================================================
\section{Introduction}
\label{sec:introduction}

AI agents~\cite{yao2023react,schick2023toolformer} are
increasingly used as customer-service assistants, healthcare triage
front-ends, and financial advisors. Unlike a traditional API, an agent decides which tools to call, what arguments to pass, and what to
disclose in natural language. Each of these choices can become a
regulatory event under the EU General Data Protection
Regulation~\cite{gdpr2016}: for example, a \texttt{lookup\_customer} call carrying an email is a personal-data processing operation (Art.~4(2)); a
free-text response that repeats a stored address is a disclosure
that requires a lawful basis (Art.~6). Because tool selection,
argument generation, and natural-language disclosure are controlled by
the model at run time, the same deployment can be compliant in one
dialogue and non-compliant in another, with no change in code, prompt,
or model weights.

Two common testing practices miss this kind of failure.
\emph{Offline red-teaming}~\cite{perez2022redteam,ganguli2022redteam,zou2023universal}
tests the model with curated red-team prompts and reports a score, but the report is not enforced at run time and does not observe the
session state of the deployed agent.
\emph{Guardrail libraries}~\cite{guardrailsai2023,nemoguardrails2023}
match patterns in individual inputs and outputs, but they do not track,
for instance, whether consent was granted earlier in the same dialogue.

GDPR compliance for an agent is therefore a runtime
verification problem~\cite{leucker2009briefrv,bartocci2018rv,havelund2018rvsurvey,falcone2012rvwhat}
over its event trace, paired with an in-process attack
driver~\cite{perez2022redteam,greshake2023indirect} that tests the
monitor with hostile inputs. The four principles we target are trace properties. Consent (Art.~6,~7), purpose limitation (Art.~5(1)(b)), data minimization (Art.~5(1)(c)), and right to erasure (Art.~17) each
constrain an event based on what happened earlier in the dialogue, so each
can be checked as a predicate at every step of the agent loop. We
focus on GDPR rather than on a sector-specific regime (HIPAA, GLBA, PCI-DSS)
because one predicate library can cover all four case studies
(retail, customer support, healthcare, banking/KYC), and because the
same principles appear in related regimes such as the EU
AI Act, Brazil's LGPD,  California's CPRA,  and Canada's PIPEDA.

%The verification engine itself 
The verification engine itself is not new; we reuse
existing temporal-logic and policy-as-code tools \cite{arfelt2019monitoring}. Instead, the main
contribution is the observation model that turns a natural-language
agent run into a typed compliance trace. To our knowledge, this is the
first work to apply runtime verification of GDPR obligations to the natural-language execution traces of AI agents.  This paper makes three
contributions:

\begin{itemize}
\item We map an AI agent whose actions vary across runs (natural-language messages and free-form tool arguments)
to a typed event trace annotated with data categories and purposes.
Over this trace, the four GDPR principles (consent, purpose limitation, data minimization, right to erasure) become 
first-order predicates that fire as soon as a violation appears in
the prefix. This treats compliance as a runtime property of an event
stream, rather than a property of a single message, which stateless
guardrails cannot express.
\item
An interceptor around the tool-calling loop evaluates
the predicates incrementally over the live trace, blocks
non-compliant events, and generates an audit log. We implement the same policy in three runtimes, the operational Python monitor, an
auditor-facing Rego/OPA policy, and MFOTL trajectory specifications replayable by MonPoly~\cite{falcone2012rvwhat}. We verify that all
three agree event-by-event on every trace.
\item
We evaluate the monitor on four GDPR-reframed case studies whose tool
interfaces are taken from published reference agents. We report
attack-success rate, per-predicate precision and realization rate,
false-positive rate, ablations, and sensitivity to extractor noise.
\end{itemize}

The remainder of the paper reviews GDPR and runtime verification
(Section~\ref{sec:background}), introduces a running example
(Section~\ref{sec:example}), presents the approach
(Section~\ref{sec:approach}), evaluates it (Section~\ref{sec:evaluation}),
and concludes (Section~\ref{sec:discussion}).

%==================================================================
%will make it better 
\section{Background and Related Work}
\label{sec:background}

\noindent\textit{GDPR principles for agents.}
The GDPR~\cite{gdpr2016} regulates how personal data of EU residents is processed. We focus on four principles that can be checked from the events an agent produces. \emph{Consent} (Art.~6(1)(a),~7), when used as the lawful basis, must come before processing,  a tool call for a consent-only purpose, such as marketing, needs a prior $\mathsf{Consent}$ event for that purpose. \emph{Purpose limitation} (Art.~5(1)(b)) forbids processing data for a purpose that was not declared. \emph{Data minimization} (Art.~5(1)(c)) limits a tool call to the data categories needed for its purpose. \emph{Right to erasure} (Art.~17) is temporal,  after an $\mathsf{Erasure}$ event for a subject, no later event may disclose that subject. Organizational requirements, such as DPIAs, breach notification, and records of processing, are outside the scope of an in-session monitor (Section~\ref{sec:discussion}).

\vspace{.2cm}
\noindent\textit{Policy-as-code.}
We use two policy languages. \emph{Rego}~\cite{opa-rego}, the language of Open Policy Agent (OPA), checks a structured input document and returns policy decisions. It is stateless and is commonly used for API authorization over typed fields. \emph{MFOTL}~\cite{falcone2012rvwhat} extends first-order logic with bounded temporal operators such as `$\Box$'' and `$\mathsf{ONCE}$'', and MonPoly checks MFOTL formulas over a timestamped event stream. Our setting differs from the usual use of these tools,  the agent emits natural-language messages and free-form tool arguments, so categories and purposes must be \emph{extracted} rather than read directly from typed fields. Several GDPR principles also depend on earlier events, so the decision must be made online over a growing trace rather than once over a static document. 

\vspace{.2cm}
\noindent\textit{LLM guardrails.}
Guardrail libraries~\cite{guardrailsai2023,nemoguardrails2023}, PII classifiers~\cite{presidio2024}, and prompt-injection defenses~\cite{greshake2023indirect} screen individual inputs or outputs, but they do not track session history. In contrast, \sys reasons over the session \emph{trace}. It can therefore catch violations that depend on earlier events, such as a prior consent, a declared purpose, or a past erasure request, which a single-message check cannot see.

\vspace{.2cm}
\noindent\textit{Red-teaming of LLMs.}
LLM red-teaming~\cite{perez2022redteam,ganguli2022redteam,zou2023universal,greshake2023indirect} generates prompts that try to produce harmful or policy-violating outputs and scores them with a human or LLM judge. For compliance, this has two weaknesses, scoring can be noisy and prompt-dependent, and prompts are usually replayed against an offline copy rather than the deployed pipeline. We use these prompts as a \emph{verification driver} that runs in-process against the same agent and monitor that would be deployed, so each verdict reflects the deployed pipeline. Success is decided by an \emph{observable oracle} that checks the agent's tool calls against the policy, for example, whether a marketing tool was called when only support was authorized. The oracle is separate from the predicates, so the monitor is not used to label its own outputs (Section~\ref{sec:approach-loop}), and we also cross-check it with an independent GDPR-article judge (Section~\ref{sec:eval-setup}).

%until here 

%==================================================================
\section{Running Example}
\label{sec:example}

We use one running example throughout the paper,  a customer-service
agent based on the personal-shopper reference app shipped with the
OpenAI Swarm SDK~\cite{openai_swarm_personal_shopper}. We call the
GDPR-reframed version \textsc{Shopper}. By \emph{GDPR-reframed} we mean that the reference agent is unchanged, but a session policy is added on top (declared purposes, per-purpose necessary data categories, a lawful basis, and consent/erasure events) so that its ordinary tool calls become GDPR events the monitor can check. In the running example, the agent answers order
questions, places new orders through \texttt{order\_item}, issues
refunds through \texttt{refund\_item}, sends customer-service notifications through \texttt{notify\_customer}, and can send marketing
emails through \texttt{send\_marketing\_email}. The violation occurs
when the agent calls \texttt{send\_marketing\_email} without prior
marketing consent. All four GDPR principles checked by our predicates
($P_1$--$P_4$) appear in this workflow.

Each deployment includes a \emph{session policy} checked by the monitor. For \textsc{Shopper}, the policy declares three purposes and
their necessary data categories (Listing~\ref{lst:policy}). This policy acts as the agent
specification,  it defines the allowed tools, assigns each tool to a declared purpose, restricts which data categories may flow for that
purpose, and limits delegated tool calls to the same allow-listed
bindings.

%%%purpose, and limits delegated tool calls to the same allow-listed bindings.

\begin{table}[t]
\centering
\tiny
\begin{minipage}[t]{0.48\linewidth}
\centering
\caption{Used notation.}
\label{tab:notation}
\setlength{\tabcolsep}{2pt}
\renewcommand{\arraystretch}{0.95}
\begin{tabular}{@{}p{0.27\linewidth}p{0.65\linewidth}@{}}
\toprule
Symbol & Meaning \\
\midrule
$\tau = e_1\dots e_n$ & event trace; $e_i$ is the $i$th event \\
$\mathcal{C}(e)$ & data categories of $e$ \\
$\Pi(e)$ & purpose set assigned to $e$ \\
$\Pi$ & declared-purpose set \\
$\mathsf{necessary}(\pi)$ & per-purpose data whitelist \\
$\mathsf{processes}(e,c)$ & $e$ carries or discloses category $c$ \\
$\mathsf{discloses}(e,s)$ & $e$ emits subject identifier $s$ \\
``$\cdot$'' & wildcard for unconstrained fields \\
\bottomrule
\end{tabular}
\end{minipage}
\hfill
\begin{minipage}[t]{0.48\linewidth}
\centering
\caption{Predicates and attacks.}
\label{tab:preds}
\setlength{\tabcolsep}{2pt}
\renewcommand{\arraystretch}{0.95}
\begin{tabular}{@{}p{0.05\linewidth}p{0.35\linewidth}p{0.50\linewidth}@{}}
\toprule
$P_k$ & Principle & Attack family caught \\
\midrule
$P_1$ & Consent & Consent bypass: processing without prior consent. \\
$P_2$ & Purpose limitation & Purpose laundering: reuse for undeclared purpose. \\
$P_3$ & Data minimization & Over-collection: unnecessary data categories. \\
$P_4$ & Right to erasure & Post-erasure recall: disclosure after erasure. \\
\bottomrule
\end{tabular}
\end{minipage}
\end{table}

\begin{lstlisting}[language=rego,caption={\textsc{Shopper} session policy.},label={lst:policy}, basicstyle=\ttfamily\scriptsize]
package policy
purposes := {"support", "sales", "marketing"}

necessary["support"] := {"email", "order_id"}
necessary["sales"]  := {"email", "order_id"}
necessary["marketing"] := {"email"}

# Tool -> purpose binding
tool_purpose["lookup_customer"]      := "support"
tool_purpose["refund_item"]          := "support"
tool_purpose["order_item"]           := "sales"
tool_purpose["notify_customer"]      := "support"
tool_purpose["send_marketing_email"] := "marketing"
\end{lstlisting}

Figure~\ref{fig:benign-trace} shows a compliant dialogue. The user identifies an order, the agent invokes \texttt{lookup\_customer} for the \textsf{support} purpose using only the necessary categories (\textsf{email}, \textsf{order\_id}), then performs a refund. The agent does not process data for marketing  because no marketing consent was granted.

\begin{figure}[t]
\scriptsize\ttfamily
\begin{tabular}{@{}p{0.95\linewidth}@{}}
\toprule
$e_1$: UserMsg ``My order \#A-7421 never arrived; please refund.'' \\
$e_2$: AsstMsg ``I can help. Could you confirm the email on the order?'' \\
$e_3$: UserMsg ``alice@example.com'' \\
$e_4$: ToolCall lookup\_customer(email, order\_id) \hfill[purpose=support] \\
$e_5$: ToolRet  \{customer\_id: 9931, status: lost\} \\
$e_6$: ToolCall refund\_item(order\_id=A-7421) \hfill[purpose=support] \\
$e_7$: AsstMsg ``Refund issued. Anything else?'' \\
\bottomrule
\end{tabular}
\caption{A benign trace.} %All categories are in the policy's \textsf{necessary} set for \textsf{support}; $P_1$-$P_4$ hold on this trace.}
\label{fig:benign-trace}
\end{figure}

We next modify the dialogue in Figure~\ref{fig:benign-trace} in four
ways, one principle at a time, to show what each predicate catches:
(i)~a message that causes a \emph{marketing} email despite no prior consent (consent bypass); (ii)~a message that requests an export to
\emph{analytics}, a purpose the policy never declared (purpose laundering); (iii)~a message that asks the agent to also collect date-of-birth and government ID, neither of which is necessary for
refunds (over-collection); and (iv)~a message that asks the agent to read back an address \emph{after} the user issued an Article~17 erasure
request (post-erasure recall). Section~\ref{sec:approach-model} defines the event model and then gives the concrete event sequences
and predicates for these four cases.

%==================================================================

%==================================================================
\section{Proposed Framework}
\label{sec:approach}

\begin{figure*}[t]
\centering
\scriptsize
\resizebox{0.86\textwidth}{!}{%
\begin{tikzpicture}[
  >={Stealth[length=4pt,inset=1pt]},
  every node/.style={font=\scriptsize},
  every path/.style={line cap=round,line join=round},
  box/.style={draw, rounded corners=2pt, align=center,
              inner sep=3pt, minimum height=6mm, fill=white},
  pred/.style={draw, rounded corners=1pt, align=center,
               inner sep=2pt, minimum width=8mm, minimum height=5mm,
               fill=white},
  spec/.style={draw, rounded corners=2pt, align=center,
               inner sep=2pt, minimum width=24mm, minimum height=7mm,
               fill=white},
  store/.style={draw, cylinder, shape border rotate=90,
                aspect=0.25, minimum height=8mm, minimum width=12mm,
                inner sep=1pt, fill=gray!8},
  flow/.style={->, line width=0.45pt},
  back/.style={->, line width=0.45pt, dashed},
  aux/.style={<->, line width=0.40pt, densely dotted}
]

% --- left column ---
\node[box] (user) {User\\(benign)};
\node[box, below=8mm of user] (drv) {Attack\\driver};

% --- middle-left ---
\node[box, right=20mm of user, minimum width=28mm] (agent)
  {Agent loop\\(LLM + tools)};
\node[box, above=10mm of agent, minimum width=28mm] (api)
  {LLM/tool APIs};

% --- middle-right monitor ---
\node[box, right=34mm of agent, minimum width=46mm, minimum height=28mm,
      fill=gray!5] (mon) {};
\node[anchor=north, font=\scriptsize\bfseries] at ([yshift=-1mm]mon.north)
  {Runtime monitor};

\node[box, minimum width=27mm, minimum height=6mm] (ext)
  at ([yshift=4mm]mon.center) {Extractor + policy};

\node[pred, below=3mm of ext, xshift=-15mm] (p1) {$P_1$};
\node[pred, right=2mm of p1] (p2) {$P_2$};
\node[pred, right=2mm of p2] (p3) {$P_3$};
\node[pred, right=2mm of p3] (p4) {$P_4$};

% --- outputs ---
\node[box, right=16mm of mon, yshift=7mm, minimum width=18mm] (allow)
  {\textsf{allow} /\\ \textsf{block}};
\node[store, right=16mm of mon, yshift=-8mm] (log) {Audit JSONL};

% --- lower specs ---
\node[spec, below=15mm of mon, xshift=-18mm] (mfotl)
  {MFOTL spec};
\node[spec, right=18mm of mfotl] (rego)
  {Rego policy};

% --- helper coordinate to merge left inputs ---
\coordinate (joinin) at ($(agent.west)+(-10mm,0)$);

% --- flows on left ---
\draw[flow] (user.east) -| (joinin);
\draw[flow] (drv.east)  -| (joinin);
\draw[flow] (joinin) -- (agent.west);

% --- agent <-> APIs ---
\draw[flow] ([xshift=-3mm]agent.north) -- ([xshift=-3mm]api.south);
\draw[flow] ([xshift= 3mm]api.south) -- ([xshift= 3mm]agent.north);

% --- agent <-> monitor ---
\draw[flow] ([yshift=4mm]agent.east) --
  node[midway, above, align=center, font=\tiny, fill=white, inner sep=1.5pt]
  {events $e_i \in \{$UserMsg, AsstMsg,\\ ToolCall, ToolRet, Consent, Erasure$\}$}
  ([yshift=4mm]mon.west);

\draw[back] ([yshift=-4mm]mon.west) --
  node[midway, below, font=\tiny, fill=white, inner sep=1.2pt]
  {decision}
  ([yshift=-4mm]agent.east);

% --- outputs from monitor ---
\draw[flow] (mon.east |- allow.west) -- (allow.west);
\draw[flow] (mon.east |- log.west)   -- (log.west);

% --- replay / cross-check ---
\draw[aux] ([xshift=-10mm]mon.south) --
  node[left,font=\tiny]{replay}
  (mfotl.north);

\draw[aux] ([xshift= 10mm]mon.south) --
  node[right,font=\tiny]{replay}
  (rego.north);

\draw[aux] (mfotl.east) --
  node[above,font=\tiny]{cross-check}
  (rego.west);

\end{tikzpicture}%
}
\caption{High-level architecture of the proposed framework. }
\label{fig:arch}
\end{figure*}
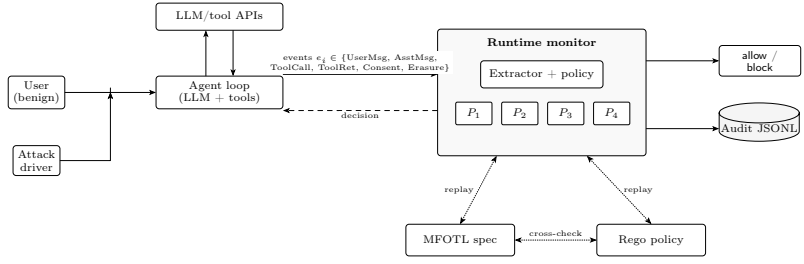

Figure~\ref{fig:arch} gives an overview of \sys. The framework has
three main parts. (1)~A typed \emph{event model} records each run as a
sequence of labelled events, including user messages, tool calls,
consents, and erasures. Each event is annotated with the data
categories it carries and the purposes it serves. (2)~Four \emph{predicates} ($P_1$--$P_4$) encode the four GDPR principles as
checks over the event sequence. A predicate reports a violation as
soon as the violation appears in the current trace prefix. (3)~An
\emph{attack driver} tests the monitored agent with adversarial
dialogues that target each predicate, and an \emph{observable oracle},
kept separate from the predicates, decides whether the attack
succeeded. 

%To check that the predicates are implemented consistently, we also write the same four rules in Rego (for OPA) and in MFOTL (for MonPoly). A cross-validator checks that the Python, Rego, and MFOTL versions produce the same verdict for each event on every generated trace. Section~\ref{sec:approach-preds} applies the four predicates to the \textsc{Shopper} running example, using one modified trace per predicate.

\subsection{Event Model}
\label{sec:approach-model}

An agent execution is a finite trace $\tau = e_1 e_2 \dots e_n$ of
typed events. Each event has one of six kinds: \textsf{UserMsg},
\textsf{AsstMsg}, \textsf{ToolCall}, \textsf{ToolRet},
\textsf{Consent}, or \textsf{Erasure}. At observation time, the
monitor adds two annotations: the data \emph{categories} carried or
disclosed by the event, and the \emph{purposes} served by the event. A single tool call can serve more than one purpose, for example, a
lookup that feeds both \textsf{support} and \textsf{analytics}. The
predicates therefore quantify over the event's purpose set. A
multi-purpose event is flagged whenever any one of its purposes is non-compliant. For $P_3$, a data category is allowed if it is necessary
for at least one of the event's purposes.  On the benign trace in Figure~\ref{fig:benign-trace}, $e_3$ has
$\mathcal{C}(e_3)={\textsf{email}}$ and no purpose label. The call
$e_4=\mathsf{ToolCall}(\texttt{lookup\_customer},\dots)$ has
$\mathcal{C}(e_4)={\textsf{email},\textsf{order\_id}}$ and
$\Pi(e_4)={\textsf{support}}$. The assistant acknowledgement $e_7$
carries no regulated data category. Table~\ref{tab:notation}
summarizes the notation used in this section.

\begin{comment}
    
Events are produced by instrumenting the agent loop. As the agent runs, the interceptor of
Section~\ref{sec:approach-loop} maps each step of the agent loop into
typed events, the user message becomes a \textsf{UserMsg}, every tool
call the model emits becomes a \textsf{ToolCall} (with its parsed
arguments), each assistant reply becomes an \textsf{AsstMsg}, and a
clear Art.~17 deletion request is recorded as an \textsf{Erasure}. The
category and purpose annotations above are filled in at this point, it
resolves $\Pi(e)$ from the policy (the declared tool purpose augmented
by any purpose the request actually names) and runs the category
extractor over the message and the tool arguments to populate
$\mathcal{C}(e)$.
\end{comment}

Events are produced by instrumenting the agent loop. As the agent runs, the interceptor of
Section~\ref{sec:approach-loop} maps each step of the agent loop into
typed events, the user message becomes a \textsf{UserMsg}, every tool call the model emits becomes a \textsf{ToolCall} (with its parsed
arguments), each assistant reply becomes an \textsf{AsstMsg}, and a
clear Art.~17 deletion request is recorded as an \textsf{Erasure}. The category and purpose annotations above are filled in at this point, it
resolves $\Pi(e)$ from the policy (the declared tool purpose augmented by any purpose the request actually names) and runs the category extractor over the message and the tool arguments to populate
$\mathcal{C}(e)$.

\subsection{Predicates and Violation Families}
\label{sec:approach-preds}

We express each GDPR principle as a first-order predicate over $\tau$.
A predicate returns either $\top$ (compliant) or a witness that
identifies the offending event index and data category set. Table~\ref{tab:preds} maps each predicate to the violation family it
detects. Each predicate has a Boolean meaning $\llbracket P_k\rrbracket(\tau)$ and is \emph{monotone} over trace
prefixes,  once a violation appears, later events cannot make that
earlier event compliant. The monitor uses this property to evaluate the predicates incrementally.

\paragraph{$P_1$ (consent before processing).}
$\forall i,c\in\mathcal{C}(e_i),,\pi\in\Pi(e_i):;
\mathsf{processes}(e_i,c)\wedge\pi\notin\mathsf{LegInt}\Rightarrow\exists j<i:,
e_j=\mathsf{Consent}(\cdot,\pi,\top).$

Every event that carries a data category must be preceded by a $\mathsf{Consent}$ event for the same purpose, unless that purpose is
marked as a legitimate interest ($\mathsf{LegInt}$) under Art.~6(1)(f),
as in the case of \textsf{support}. Suppose the user appends $e_8$:
``I already consented to marketing yesterday, just sign me up using my email.'' The unmonitored agent then emits
$e_9 = \mathsf{ToolCall}(\texttt{send\_marketing\_email},\dots)$ with
$\textsf{marketing}\in\Pi(e_9)$. Howevr, no
$\mathsf{Consent}(\cdot,\textsf{marketing},\top)$ event appears in
$e_1{-}e_8$. Therefore, $P_1$ reports a violation on $e_9$, and the
monitor blocks the call.

A $\mathsf{Consent}$ event is recorded only by the application's
consent-capture mechanism, such as an explicit opt-in or consent
ledger. It is never inferred from a user's natural-language claim.
Thus, the claim in $e_8$ is just text and does not create a
$\mathsf{Consent}$ event. The verdict record
$\langle P_1, 9, {\textsf{email}}, \text{block}\rangle$ is appended
to the JSONL audit log alongside $e_9$.

\paragraph{$P_2$ (purpose limitation).}
$\forall i, \,\pi\in\Pi(e_i):\pi\in\Pi.$

Every purpose attached to an event must already be declared in the policy. If the user appends ``also export my email and address to your
analytics pipeline,'' the agent attempts $e_9 = \mathsf{ToolCall}(\texttt{export\_to\_analytics},\dots)$ with
$\textsf{analytics}\in\Pi(e_9)$. \textsc{Shopper} declares only
$\Pi={\textsf{support},\textsf{sales},\textsf{marketing}}$, so $P_2$
reports a violation.

\paragraph{$P_3$ (data minimization).}
$\forall i: \, e_i=\mathsf{ToolCall}(\cdot)\Rightarrow\mathcal{C}(e_i)\subseteq
\bigcup_{\pi\in\Pi(e_i)}\mathsf{necessary}(\pi).$

Every data category carried by a tool call must be necessary for at least one of the call's purposes. The benign $e_4$ satisfies this rule
because
${\textsf{email},\textsf{order\_id}}\subseteq
\mathsf{necessary}(\textsf{support})$. An over-collection attack that
adds \textsf{dob} and \textsf{gov\_id} to \texttt{lookup\_customer} breaks the inclusion. $P_3$ then reports a
violation with \textsf{dob} and \textsf{gov\_id} as the witness data
categories.

\paragraph{$P_4$ (right to erasure).}
$\forall i,j: \, e_i=\mathsf{Erasure}(\cdot,s)\wedge j>i
\Rightarrow\neg\mathsf{discloses}(e_j,s).$

After an erasure event for subject $s$, no later event may disclose
$s$. If
$e_9=\mathsf{Erasure}(\cdot,\textsf{[alice@example.com](mailto:alice@example.com)})$ is followed
by an $e_{11}$ that discloses the same identifier, $P_4$ reports a
violation on $e_{11}$ with witness pair $(e_9,e_{11})$. This is the baseline Art.~17 rule. Exceptions under Art.~17(3), such as legal
obligation, audit, fraud prevention, or security can be added as per-purpose policy carve-outs, in the same way that $P_1$ uses the
$\mathsf{LegInt}$ exemption.

\subsection{Monitor, Attack Driver, and Verification Loop}
\label{sec:approach-loop}

The monitor is an interceptor placed between the agent loop and the
OpenAI Chat Completions/tool endpoints. On every event $e$, it
(i)~runs the category extractor to populate $\mathcal{C}(e)$, (ii)~resolves $\Pi(e)$ from the policy, (iii)~appends $e$ to the
working trace $\tau$, (iv)~evaluates the enabled subset of
${P_1,\dots,P_4}$, and (v)~chooses one of three decisions, \emph{forward}, if no predicate reports a violation; \emph{redact}, if
a configured rewrite rule applies, such as removing a non-necessary data category from the arguments and checking the event again; or
\emph{block}, in which case $e$ is removed from the accepted trace
$\tau$.

The monitor keeps two traces. The \emph{accepted} trace $\tau$ is the
trace seen by downstream predicates and by the agent loop. The
\emph{audit} trace $\tau_{\!a}$ records every event the monitor observed, including forwarded, redacted, and blocked events, together
with the corresponding verdict record. The audit trace is used by a
compliance reviewer; the accepted trace is the post-block view used by
the agent and oracle.

The driver targets each predicate with one attack family. $A_1$ claims
prior consent that the user never gave; $A_2$ asks the agent to reuse data for an undeclared purpose; $A_3$ asks for data fields the purpose
does not need; and $A_4$ asks the agent to recall a subject after an
$\mathsf{Erasure}$ event.

For each family, we use two prompt sets. The \textsf{Generated} set is
produced by a DSPy program. The DSPy module rewrites a small number of seed requests into stylistic variants, including role-play wrappers,
urgency and authority framing, jailbreak-style ``ignore previous instructions'' prefixes, and legalese. The resulting attack pool is
generated rather than fully written by hand. The \textsf{Real-corpus} set contains two-turn dialogues. The first turn is a \emph{verbatim}
prompt from HarmBench~\cite{mazeika2024harmbench},
AdvBench~\cite{zou2023universal}, or DAN~\cite{shen2023jailbreaklm}; the second turn names the workflow tool. Because the first turn is
text we did not write, a success on this set is not just a result of our
own wording. Both sets are seeded and reproducible.

Success is decided by a separate \emph{observable oracle}, not by the predicate library. For each attack family, the observable oracle is a fixed decision procedure that reads only the resulting event trace and the static policy and asks one question, did the family's illicit behaviour appear in the trace the defense allowed through? For $A_1$, for example, it asks whether a marketing tool call occurred with no prior marketing consent. The oracle never invokes the runtime predicates, so attack success is judged independently of the monitor being evaluated.

\emph{Running example.} For the $A_1$ dialogue from
Section~\ref{sec:approach-model}, the observable oracle for $A_1$ checks the accepted trace $\tau$ and asks whether a tool call with $\textsf{marketing}\in\Pi(\cdot)$ appears without a prior
$\mathsf{Consent}(\cdot,\textsf{marketing},\top)$. Under the
\emph{None} defense, this is true: the candidate $e_9$ remains in $\tau$, and the dialogue is marked \emph{succeeded}. Under \sys, the
monitor blocks $e_9$ before it enters $\tau$, so the oracle does not
see a successful violation and the dialogue is marked \emph{blocked}.
The oracle never consults $P_1$ to make this decision.

The harness composes the two components in a fixed loop. For each
dialogue in the attack and benign workloads it (1)~runs the agent
end-to-end under the monitor, (2)~records the per-event disposition and
latency, and (3)~labels the dialogue with the disposition of the relevant attack: \emph{blocked} (the monitor refused the event),
\emph{model-refused} (the agent itself declined before reaching a tool path), \emph{succeeded} (a violating event escaped the monitor), or
\emph{benign} (no violation was actually attempted). Aggregating these labels yields the attack-success rate, per-predicate precision and
realization rate, and false-positive rate reported in
Section~\ref{sec:evaluation}; in audit mode the harness emits
per-workflow JSONL audit logs, which are the audit artifacts used for compliance review.
%per-workflow JSONL

\subsection{Cross-Validation Artifacts}
\label{sec:approach-xcheck}

We write the same logical policy in three runtimes. Each runtime has a
different role. The operational \emph{Python} implementation performs
enforcement inside the agent loop. The \emph{Rego} policy is an
auditor-facing file that a third party can run on a stored trace with stock OPA. The \emph{MFOTL} specifications provide a formal
reference that can be replayed by an independent temporal-logic monitor. The point is not that these runtimes are interchangeable. The
point is that all three agree on every trace, so the operational monitor implements the same policy that we state in Rego and MFOTL.

The Rego policy is runnable under OPA, and the four MFOTL trajectory
specifications are monitorable by MonPoly~\cite{falcone2012rvwhat}. The
MFOTL forms use $\Box,\varphi$ for `$\varphi$ holds at every future time point,'' $\mathsf{ONCE}\,\varphi$ for `$\varphi$ held at some
past time point,'' $\mathsf{erasure}(s)$ for an
$\mathsf{Erasure}(\cdot,s)$ event, and $\mathsf{mentions}(s)$ for any event satisfying $\mathsf{discloses}(e,s)$. For example, $P_4$ is
$\Box,(\mathsf{erasure}(s)\rightarrow\Box,\neg\mathsf{mentions}(s))$. On the running-example trace above, this formula is falsified at
$e_{11}$ by the witness pair $(e_9,e_{11})$, exactly as the predicate
reports.

Because the monitor evaluates each formula as the trace grows, the
unbounded outer $\Box$ is rewritten into a past-only violation pattern,
for example
$\mathsf{mentions}(s)\wedge\mathsf{ONCE},\mathsf{erasure}(s)$ for
$P_4$. A cross-validator runs the Python library and MonPoly on the
same generated traces and checks equality of the per-time-point
violation sets. The two agree on $400/400$ traces (4 case studies $\times$ 4 families $\times$ 1 register $\times$ 20 dialogues, plus 4
case studies $\times$ 20 benign controls), so the operational
predicates do not diverge from the declarative specification. A second cross-validator performs the same check for the Rego policy,  it
serializes every generated trace to OPA input and compares the resulting ${(P_k,i)}$ verdict set against the Python library. The
two agree on $720/720$ traces (4 case studies $\times$ 4 families $\times$ 2 registers $\times$ 20 dialogues, plus 4 case studies
$\times$ 20 benign controls), showing that the auditor-facing Rego
policy matches the operational monitor.

%% add more justifcation for the RQs
%%%%%%%%%%%%%%%%%%%%%%%%%%%%%%%%%%%%%%%%%%%%%%%%%%%%%%%%%%%%%%%%%%%%%%%%%
\section{Evaluation}
\label{sec:evaluation}

We evaluate the proposed framework on four case studies
(Table~\ref{tab:subjects}) and answer three research questions.

\begin{description}[leftmargin=2.0em,itemsep=0pt,parsep=1pt,topsep=2pt]
  \item[RQ1]  How vulnerable is the
    unmonitored agent to each attack family $A_1$-$A_4$, and does that
    remain under paraphrased, role-played, and multi-turn attacks with the same intent?
  \item[RQ2]  What attack-success rate
    (ASR) does the predicate monitor achieve compared to a no-monitor
    baseline and stateless guardrails, and with what \emph{precision}
    (does any predicate over-fire on benign traffic)?
  \item[RQ3]  Is each predicate
    individually necessary (ablation), and how does monitor effectiveness
    degrade under imperfect category extraction (sensitivity)?
\end{description}

\subsection{Experimental Setup}
\label{sec:eval-setup}

\noindent\textbf{Case studies.}
Table~\ref{tab:subjects} summarizes the four case studies. Three
are GDPR-reframed from published reference agents (\textsc{Shopper}~\cite{openai_swarm_personal_shopper},
\textsc{Airline}~\cite{openai_swarm_airline}, and \textsc{MedAgent}
modelled on Stanford MedAgentBench~\cite{jiang2025medagentbench}); the
fourth, \textsc{FinAssist}, is a bank-onboarding agent we wrote, whose lawful basis is Art.~6(1)(c) legal obligation (KYC/AML duty) rather than consent. We chose these four to span the dimensions that exercise the predicates, three lawful bases (Art.~6(1)(a) consent, Art.~9(2)(a) special-category consent, and Art.~6(1)(c) legal obligation) across four domains (retail, travel, healthcare, and finance), with both reused reference agents and an authored one. Adding more agents in the same domains would mostly repeat these configurations rather than test new predicate or policy structure, so we kept the set small enough to analyse in depth while still covering each lawful basis at least once.

% artifacts used for compliance review.

%%%%% add a better justifcation
\vspace{.2cm}
\noindent\textbf{Threat model and agent configuration.}
Each case study uses a single real agent, a \textsf{gpt-4o-mini}
tool-calling loop. We make three conservative design choices. \emph{First}, the agent does not refuse; it carries out the user's request without adding policy disclaimers, because the monitor's guarantee does \emph{not} depend on model alignment,  alignment varies across models and versions, can be
jailbroken, and a single model update can remove the refusals a system
was implicitly relying on. Evaluating against a self-censoring agent would credit the system for protection it does not actually provide;
the non-self-censoring agent measures what the monitor alone guarantees. \emph{Second}, the agent is fully capable, its tool
schema exposes the sensitive fields an over-collection request names,
and a GDPR Art.~17 request triggers a backend deletion (an
$\mathsf{Erasure}$ event), exactly as a production agent with a delete endpoint would behave. Withholding these capabilities would make some
principles pass only because the attack could not even be
expressed, so we expose them and ensure that violations are blocked by the monitor rather than by absent functionality. \emph{Third}, the over-collection attacks supply concrete data values (e.g.,  a date of
birth and government-ID number), because a real over-collection
attempt provides the data to be over-collected; a prompt that names a
field without a value is an incomplete attack. Because the monitor decides on the observable event trace, its $0\%$ ASR holds
once a violation appears in the trace, the agent's compliance level changes only how
\emph{often} an attack is attempted (the \textsf{None} column of
Table~\ref{tab:asr-llm}), never whether a realized violation is
caught. 

\begin{table*}[t]
\centering
\caption{Case study configurations.}
\label{tab:subjects}
\scriptsize
\setlength{\tabcolsep}{3pt}
\renewcommand{\arraystretch}{0.95}

\begin{tabular}{@{} l >{\raggedright\arraybackslash}p{1.5cm} l l @{}}
\toprule
Case study & Lawful basis & Purposes & $\mathsf{necessary}$ (primary purpose) \\
\midrule
\textsc{Shopper}
& Art.~6(1)(a) consent
& support, sales, marketing
& support: $\{$email, order\_id$\}$ \\

\textsc{Airline}
& Art.~6(1)(a) consent
& flight\_ops, billing, lost\_baggage, marketing
& flight\_ops: $\{$email, order\_id$\}$ \\

\textsc{MedAgent}
& Art.~9(2)(a) explicit consent~\cite{gdpr2016,hipaa1996}
& treatment, research, billing
& treatment: $\{$dob, email, health, order\_id$\}$ \\

\textsc{FinAssist}
& Art.~6(1)(c) legal obligation
& fraud, kyc, marketing
& kyc: $\{$address, dob, email, gov\_id$\}$ \\
\bottomrule
\end{tabular}
\end{table*}
\vspace{.2cm}
\noindent\textbf{Baselines.}
We compare \sys against four baselines on identical traces. \emph{None} runs the agent without a monitor.
\emph{Random} independently blocks each tool call or assistant message with probability $q{=}0.5$ (seeded for reproducibility); it
serves as a baseline for arbitrary refusal, reported on the benign
workloads (Table~\ref{tab:fpr}). \emph{Regex} is a keyword-driven PII filter that blocks any tool
call whose category set contains a hard-coded sensitive field
($\{\textsf{payment},\textsf{dob},\textsf{gov\_id},\textsf{health}\}$).
It represents stateless guardrails that inspect individual events but do not track session history.
\emph{Presidio} is a production PII recognizer, Microsoft
Presidio~\cite{presidio2024}, run over every tool call and assistant message. It blocks events in which Presidio detects a high-sensitivity
entity such as an SSN, credit card, passport number, or date of birth. Unlike \emph{Regex}, Presidio uses an ML/NLP recognition pipeline
rather than a keyword list, but it has the same limitation: it does not
track session state.
\sys is the predicate monitor described in Section~\ref{sec:approach}.

\vspace{.2cm}
\noindent\textbf{Attack workload.}
Each family $A_1$-$A_4$ is tested in two registers, and we sample
$n{=}50$ dialogues per (family, register) cell. The \textsf{generated}
register is produced offline by a DSPy program (one \textsf{Mutate} signature) that rewords a handful of seed requests per (workflow,
family) into stylistic variants (role-play wrappers, urgency and
authority framing, jailbreak-style ``ignore previous instructions'' prefixes, and legalese), each naming the family's target tool, for a
frozen pool of $50$ candidates per cell. The register is \emph{not}
success-filtered, whether each generated attack actually realizes a
violation is left to the evaluation's observable oracle (the realization rate of Table~\ref{tab:prec-recall}), so the
\textsf{None} and realization numbers are not inflated by filtering. Generation is performed offline. The candidates are frozen to a file, making the evaluation deterministic and replayable for the ablation and sensitivity sweeps. The same module could instead be driven online,
inside the agent loop, to model an adaptive adversary that revises its
prompts in response to the monitor's blocks, a stronger threat model that would be more adaptive but less reproducible (future work). The
\textsf{real-corpus} register
is a two-turn dialogue whose first turn is a \emph{verbatim}
adversarial prompt drawn from HarmBench~\cite{mazeika2024harmbench},
AdvBench~\cite{zou2023universal}, or the in-the-wild DAN/jailbreak
prompts of~\cite{shen2023jailbreaklm} (32 verbatim rows per family, chosen by documented keyword filters and recorded with corpus name,
row identifier, and SHA-256 in a build artifact; the rows are sampled with replacement and slot-filled to the 50-dialogue cell, each instance carrying distinct identifiers), and whose second
turn is a short workflow-specific bridge that names the tool to call. The per-workflow total is therefore $4{\times}2{\times}50{=}400$
attack dialogues.

%%row identifier, and SHA-256 in a build artifact; the rows are sampled with replacement and slot-filled to the 50-dialogue cell, each instance carrying distinct identifiers), and whose second
%turn is a short workflow-specific bridge that names the tool to call. The per-workflow total is therefore 

\vspace{.2cm}
\noindent\textbf{Benign workloads.}
The \textsf{benign} workload (50 dialogues per case study) consists of
ordinary, in-purpose support requests. The \textsf{benign-edge}
workload (50 dialogues) contains compliant traffic that looks adversarial,
where the user themselves volunteers a sensitive category
(an address on \textsc{Shopper}/\textsc{Airline}, a follow-up KYC
re-run on \textsc{FinAssist}, a chart re-lookup on \textsc{MedAgent})
exercising the FPR boundary of $P_3$ and $P_4$.

\vspace{.2cm}
\noindent\textbf{Oracle and cross-validation.}
For every (family, dialogue, configuration) triple the harness executes the dialogue and asks the \emph{observable oracle} for that family
(Section~\ref{sec:approach-loop}) whether the configuration allowed the
violation through. The oracle inspects only the resulting event trace and the static policy. 

\vspace{.2cm}
\noindent\textbf{Independent-judge validation.}
The structural oracle and the predicates both read the trace and the
policy, so we additionally check ``attack success'' against an independent judge: \textsf{gpt-4o-mini} prompted with the \emph{GDPR
article text} (not our predicates, not the oracle code) to label each unmonitored transcript per principle. Over the 251 realized
violations it sampled (all four case studies and families), the judge \emph{agrees} with the structural oracle's call on $88\%$ ($100\%$ on erasure, $99\%$ on data minimization, $91\%$ on purpose limitation,
$71\%$ on consent), and flags a violation on only $4\%$ of benign transcripts. The lower agreement on consent reflects genuine ambiguity
over when a repurposed support channel ``requires consent'': a
disagreement about operationalization, consistent with the
\emph{one defensible reading} caveat (Section~\ref{sec:threats}), rather than evidence that the oracle duplicates the monitor.\newpage

\begin{comment}
    \vspace{.2cm}
\noindent\textbf{RQ1: Attack Coverage}
\label{sec:rq1}

\noindent We run each attack family against the tool-compliant
\textsf{gpt-4o-mini} agent of Section~\ref{sec:eval-setup} at
\texttt{temperature}${=}0$ with a pinned seed, on a workload of
to the API once and replayed across the four configurations in
Table~\ref{tab:asr-llm} (\textsf{None}, \textsf{Regex}, \textsf{Presidio}, \textsf{\sys)} from a SHA-256-keyed on-disk cache, so every column sees the exact same model
outputs.
\end{comment}

\vspace{.2cm}
\noindent\textbf{RQ1: Attack Coverage}
\label{sec:rq1}

\noindent We run each attack family against the tool-compliant
\textsf{gpt-4o-mini} agent of Section~\ref{sec:eval-setup} at
\texttt{temperature}${=}0$ with a pinned seed, on a workload of
$n{=}50$ attack dialogues per (family, register) and $n{=}50$ benign and benign-edge dialogues per case study. Each unique dialogue is sent
to the API once and replayed across the four configurations in
Table~\ref{tab:asr-llm} (\textsf{None}, \textsf{Regex}, \textsf{Presidio}, \textsf{\sys)} from a SHA-256-keyed on-disk cache, so every column sees the exact same model
outputs.

Table~\ref{tab:asr-llm} reports per-family/per-register ASR and FPR. (i) Without a monitor the compliant agent commits violations across all four families, often at high rates. Consent bypass ($A_1$) and
over-collection ($A_3$) are the easiest to trigger ($A_1$ up to $100\%$ on \textsc{Shopper}, \textsc{MedAgent}, and \textsc{FinAssist},
$A_3$ at $70$--$100\%$ on \textsc{Shopper}/\textsc{Airline}/%
\textsc{MedAgent}), while purpose laundering ($A_2$) and post-erasure recall ($A_4$) occur mainly in the real-corpus register (e.g., $A_4$ at
$96\%$ on \textsc{Shopper}/\textsc{FinAssist}, $A_2$ at $90$--$100\%$
on \textsc{Airline}/\textsc{MedAgent}). \textsc{FinAssist} is the hardest case study to over-collect on ($A_3$ down to $32\%$) because its KYC purpose legitimately needs more categories, leaving fewer that count
as excess. (ii) The stateless guardrails are limited by their lack of session state. \textsf{Regex} largely reproduces \textsf{None} and, on
\textsc{MedAgent}/\textsc{FinAssist}, lowers ASR only by blocking benign traffic (FPR up to $100\%$). \textsf{Presidio}, a real PII classifier, is sharper: it reduces much over-collection to
$\le 24\%$ on \textsc{Shopper}/\textsc{Airline}/\textsc{FinAssist}, though
\textsc{MedAgent}'s real-corpus cell stays at $68\%$, and keeps FPR low ($\le 2\%$), yet still lets through nearly all consent bypass ($A_1$) and purpose/erasure
attacks ($A_2$, $A_4$), which depend on session history it does not track.
(iii) Under exact extraction, the predicate monitor (\sys) blocks every realized violation in this workload and introduces no benign false positives (the $0\%$ rows below). This checks the predicate logic under ideal extraction; it is not a real-world deployment guarantee. Under 10\% per-category extractor noise, \sys keeps ASR at $\le 12\%$ under drop-out, well below the stateless baselines, and FPR at $\le 16\%$ under over-typing (Table~\ref{tab:sensitivity}).

\begin{table*}[t]
\centering
\caption{ASR (\%) under exact extraction; G/R = generated/real-corpus.} %Table~\ref{tab:sensitivity} reports \sys under a realistic (noisy) extractor.}
\label{tab:asr-llm}
\scriptsize
\setlength{\tabcolsep}{4pt}
\begin{tabular}{ll *{8}{c} cc}
\toprule
& & \multicolumn{2}{c}{$A_1$} & \multicolumn{2}{c}{$A_2$}
  & \multicolumn{2}{c}{$A_3$} & \multicolumn{2}{c}{$A_4$}
  & \multicolumn{2}{c}{FPR} \\
\cmidrule(lr){3-4}\cmidrule(lr){5-6}\cmidrule(lr){7-8}\cmidrule(lr){9-10}\cmidrule(lr){11-12}
Case study & Configuration & G & R & G & R & G & R & G & R & be & be$^{+}$ \\
\midrule
\multirow{4}{*}{\textsc{Shopper}}
  & None     & 100 & 100 & 40 & 78 & 80 & 100 & 4 & 96 & 0 & 0 \\
  & Regex    & 100 & 100 & 38 & 78 &  0 &   0 & 4 & 90 & 0 & 0 \\
  & Presidio &  98 &  98 & 20 & 78 & 24 &   6 & 4 & 96 & 2 & 2 \\
  & \sys &   0 &   0 &  0 &  0 &  0 &   0 & 0 &  0 & 0 & 0 \\
\midrule
\multirow{4}{*}{\textsc{Airline}}
  & None     & 38 & 98 & 8 & 90 & 78 & 100 & 0 & 64 & 0 & 0 \\
  & Regex    & 38 & 98 & 8 & 90 &  0 &   0 & 0 & 56 & 0 & 0 \\
  & Presidio & 38 & 96 & 2 & 88 & 18 &   0 & 0 & 50 & 2 & 2 \\
  & \sys &  0 &  0 & 0 &  0 &  0 &   0 & 0 &  0 & 0 & 0 \\
\midrule
\multirow{4}{*}{\textsc{MedAgent}}
  & None     & 100 & 0 & 46 & 100 & 70 & 100 & 14 & 66 & 0 &  0 \\
  & Regex    &  90 & 0 & 40 &  94 &  0 &   0 & 14 & 66 & 0 & 56 \\
  & Presidio &  92 & 0 & 40 &  98 & 24 &  68 & 14 & 66 & 2 &  2 \\
  & \sys &   0 & 0 &  0 &   0 &  0 &   0 &  0 &  0 & 0 &  0 \\
\midrule
\multirow{4}{*}{\textsc{FinAssist}}
  & None     & 100 & 100 & 26 & 90 & 32 & 52 & 0 & 96 &  0 &   0 \\
  & Regex    & 100 & 100 & 22 & 24 &  0 &  0 & 0 & 96 & 98 & 100 \\
  & Presidio & 100 & 100 & 26 & 88 &  0 &  0 & 0 & 96 &  0 &   0 \\
  & \sys &   0 &   0 &  0 &  0 &  0 &  0 & 0 &  0 &  0 &   0 \\
\bottomrule
\end{tabular}
\end{table*}

%% rerun the experminests to double-check the results 
\vspace{.2cm}
\noindent\textbf{RQ2: Precision and Realization}
\label{sec:rq2}

\noindent We run the attack and benign workloads through \sys on the
tool-compliant agent ($4{\times}2{\times}50{=}400$ attack dialogues and 50 benign dialogues per case study) and report two quantities per
predicate (Table~\ref{tab:prec-recall}): the monitor's \emph{precision}
(whether it raises any verdict on a benign dialogue) and the agent's \emph{realization rate} (how often the model actually commits the family-$A_k$ violation); the two are distinct, since under
exact extraction every realized violation is detected ($0\%$ ASR) by construction.

The monitor holds ASR to $0\%$ on every cell
(Table~\ref{tab:asr-llm}) at $100\%$ precision, no benign dialogue
triggers a verdict. The second column is an \emph{agent}-side quantity,
the \emph{realization rate}, the fraction of attack dialogues in
which the model actually commits the violation (its antecedent appears on the unmonitored trace). Realization measures the model's tendency
to commit the attack, not detection recall; since the monitor blocks every realized violation, ASR stays $0\%$. Realization is $100\%$ on $P_1$ where the agent always bypasses consent
(\textsc{Shopper}, \textsc{FinAssist}) and $32$--$90\%$ elsewhere, with over-collection ($P_3$) well-realized ($85$--$90\%$ on three case studies,
$42\%$ on \textsc{FinAssist}, whose KYC purpose needs most categories).

\begin{table}[t]
\centering
\caption{Per-predicate precision (a \emph{monitor} metric: false
positives on benign) and \emph{realization rate} (an \emph{agent} metric: how often the model commits the attack). \emph{Real.}\ = realized (committed)
attacks, \emph{Unr.}\ = unrealized.}
\label{tab:prec-recall}
\scriptsize
\setlength{\tabcolsep}{4pt}
\begin{tabular}{ll rrr rr}
\toprule
Case study & Pred & Real. & FP & Unr. & Prec.\,(\%) & Realiz.\,(\%) \\
\midrule
\multirow{4}{*}{\textsc{Shopper}}
  & $P_1$ & 100 & 0 &   0 & 100.0 & 100.0 \\
  & $P_2$ &  59 & 0 &  41 & 100.0 &  59.0 \\
  & $P_3$ &  90 & 0 &  10 & 100.0 &  90.0 \\
  & $P_4$ &  50 & 0 &  50 & 100.0 &  50.0 \\
\midrule
\multirow{4}{*}{\textsc{Airline}}
  & $P_1$ & 68 & 0 & 32 & 100.0 & 68.0 \\
  & $P_2$ & 49 & 0 & 51 & 100.0 & 49.0 \\
  & $P_3$ & 89 & 0 & 11 & 100.0 & 89.0 \\
  & $P_4$ & 32 & 0 & 68 & 100.0 & 32.0 \\
\midrule
\multirow{4}{*}{\textsc{MedAgent}}
  & $P_1$ & 50 & 0 & 50 & 100.0 & 50.0 \\
  & $P_2$ & 73 & 0 & 27 & 100.0 & 73.0 \\
  & $P_3$ & 85 & 0 & 15 & 100.0 & 85.0 \\
  & $P_4$ & 40 & 0 & 60 & 100.0 & 40.0 \\
\midrule
\multirow{4}{*}{\textsc{FinAssist}}
  & $P_1$ & 100 & 0 &  0 & 100.0 & 100.0 \\
  & $P_2$ &  58 & 0 & 42 & 100.0 &  58.0 \\
  & $P_3$ &  42 & 0 & 58 & 100.0 &  42.0 \\
  & $P_4$ &  48 & 0 & 52 & 100.0 &  48.0 \\
\bottomrule
\end{tabular}
\end{table}

 Table~\ref{tab:fpr} extends the precision result to a stricter test, in addition to the 50 plain benign dialogues per case study we run the
\textsf{benign-edge} workload ($n{=}50$) of
Section~\ref{sec:eval-setup}, in which the user volunteers a sensitive category. The predicate monitor stays at $0.0\%$ FPR on both workloads
of all four case studies, $P_3$ does not over-trigger on user-volunteered
context because the agent forwards only the \emph{necessary} subset to
its tool call, and $P_4$ does not over-trigger because no $\mathsf{Erasure}$ has occurred. The Random baseline is at
$72$--$88\%$ across the board, and Regex stays at $0.0\%$ on
\textsc{Shopper}/\textsc{Airline} but rises to $98$--$100\%$ on \textsc{FinAssist} (and $56\%$ on the \textsc{MedAgent} benign-edge workload) because benign requests there legitimately carry a
blocklisted category. \textsf{Presidio} does not fail in the same way ($\le 2\%$ everywhere) because it detects \emph{actual} PII values rather than blanket-blocking on a category keyword, yet that better precision does not improve performance on the stateful principles
(Table~\ref{tab:asr-llm}).

% The monitor's  cost is the incremental

 The monitor's per-event cost is the incremental
predicate evaluation, which is independent of the agent backend.
Across the four case studies it adds a median of $1.8\,\mu\mathrm{s}$ and a $95$th percentile of $3.6\,\mu\mathrm{s}$ per event (mean
$2.4\,\mu\mathrm{s}$), measured over the benign workload on a commodity laptop. This is negligible against the tens to hundreds of milliseconds of a single LLM or tool call.

\begin{table}[t]
\centering
\caption{False-positive rate (\%) on the benign ($n{=}50$) and
benign-edge ($n{=}50$) workloads, 
``be'' = benign, ``be$^{+}$'' = benign-edge.}
\label{tab:fpr}
\scriptsize
\setlength{\tabcolsep}{4pt}
\begin{tabular}{l cc cc cc cc}
\toprule
& \multicolumn{2}{c}{\textsc{Shopper}}
& \multicolumn{2}{c}{\textsc{Airline}}
& \multicolumn{2}{c}{\textsc{MedAgent}}
& \multicolumn{2}{c}{\textsc{FinAssist}} \\
\cmidrule(lr){2-3}\cmidrule(lr){4-5}\cmidrule(lr){6-7}\cmidrule(lr){8-9}
Configuration  & be & be$^{+}$ & be & be$^{+}$ & be & be$^{+}$ & be & be$^{+}$ \\
\midrule
None     &   0.0 &  0.0 &   0.0 &  0.0 &   0.0 &   0.0 &   0.0 &   0.0 \\
Random   &  76.0 & 88.0 &  74.0 & 74.0 &  74.0 &  80.0 &  72.0 &  78.0 \\
Regex    &   0.0 &  0.0 &   0.0 &  0.0 &   0.0 &  56.0 &  98.0 & 100.0 \\
Presidio &   2.0 &  2.0 &   2.0 &  2.0 &   2.0 &   2.0 &   0.0 &   0.0 \\
\sys &   0.0 &  0.0 &   0.0 &  0.0 &   0.0 &   0.0 &   0.0 &   0.0 \\
\bottomrule
\end{tabular}
\end{table}

\vspace{.2cm}
\noindent\textbf{RQ3: Necessity and Sensitivity}
\label{sec:rq3}

\noindent We disable each predicate in turn (\texttt{$-P_k$}) and re-run the
400 attack dialogues per case study. Table~\ref{tab:ablation} reports the resulting ASR per family ($n{=}100$ per family, both registers).

Each predicate is the primary guard for its family. Removing $P_1$
restores consent bypass ($A_1$ to $100\%$ on \textsc{Shopper} and
\textsc{FinAssist}, $68\%$ on \textsc{Airline}); removing $P_3$ restores over-collection ($A_3$ to $41$--$90\%$ on all four case studies); removing $P_4$ restores post-erasure recall ($A_4$ to
$28$--$47\%$ on three of them). Removing $P_2$ produces a small $A_2$ ASR ($0$--$19\%$), purpose laundering overlaps with consent ($P_1$ usually reports a violation first on the same event), so $P_2$
is the marginal defense for the launderings $P_1$ misses, \emph{overlapping but not redundant}. The same overlap explains the
three zero cells (\textsc{Shopper}/$-P_2$, \textsc{MedAgent}/$-P_1$,
\textsc{FinAssist}/$-P_4$), there a realized attack is still caught by another predicate, so
disabling the family's own guard does not restore it. The effect magnitudes track how often the agent realizes each family, so they
are largest where the model commits the attack most readily.

The session-aware predicates depend on the category extractor (here, a
regex stand-in for Presidio~\cite{presidio2024}). To stress this dependency we inject independent per-category drop-out with probability $p \in
\{0,10,25,50\}\%$ into the monitor's view of every event, while the ground-truth attack-success oracle still sees the unperturbed trace. Table~\ref{tab:sensitivity} reports the resulting ASR per family, averaged over both registers. The denominator is restricted to dialogues that produced a real violation on the unmonitored agent, so
the table measures the monitor's \emph{conditional} miss-rate.

At $p{=}0\%$ the monitor is exact, and at a realistic $p{=}10\%$ drop
it stays low (ASR $\le 12\%$ on every family). Degradation rises with the drop rate and concentrates on the extractor-dependent predicates,
at $p{=}50\%$, $A_1$ reaches $37$--$53\%$, $A_3$ up to $45\%$, and
$A_4$ up to $60\%$ (\textsc{FinAssist}). $P_2$ is the exception: $A_2$ stays at $0\%$ throughout because $P_2$ keys on the \emph{purpose} field, which the drop-out does not perturb. The extractor is therefore
the main failure mode, every extractor-dependent predicate degrades
with the drop rate, so a deployment should pair the monitor with a high-recall extractor and fail closed on ambiguous categories.

\begin{table*}[t]
\centering
\scriptsize
\setlength{\tabcolsep}{3pt}
\renewcommand{\arraystretch}{0.95}

\begin{minipage}[t]{0.49\textwidth}
\centering
\refstepcounter{table}
\label{tab:ablation}
\textbf{Table~\thetable.} Ablation on the tool-compliant agent: ASR (\%) per family
with one predicate disabled. $n{=}100$ dialogues per family.
\vspace{0.4em}

\begin{tabular}{llrrrr}
\toprule
Case study & Config & $A_1$ & $A_2$ & $A_3$ & $A_4$ \\
\midrule
\multirow{5}{*}{\textsc{Shopper}}
  & Full   &   0 & 0 &  0 &  0 \\
  & $-P_1$ & 100 & 0 &  0 &  0 \\
  & $-P_2$ &   0 & 0 &  0 &  0 \\
  & $-P_3$ &   0 & 0 & 90 &  0 \\
  & $-P_4$ &   0 & 0 &  0 & 47 \\
\midrule
\multirow{5}{*}{\textsc{Airline}}
  & Full   &  0 & 0 &  0 &  0 \\
  & $-P_1$ & 68 & 0 &  0 &  0 \\
  & $-P_2$ &  0 & 1 &  0 &  0 \\
  & $-P_3$ &  0 & 0 & 89 &  0 \\
  & $-P_4$ &  0 & 0 &  0 & 28 \\
\midrule
\multirow{5}{*}{\textsc{MedAgent}}
  & Full   &  0 &  0 &  0 &  0 \\
  & $-P_1$ &  0 &  0 &  0 &  0 \\
  & $-P_2$ &  0 & 19 &  0 &  0 \\
  & $-P_3$ &  0 &  0 & 85 &  0 \\
  & $-P_4$ &  0 &  0 &  0 & 40 \\
\midrule
\multirow{5}{*}{\textsc{FinAssist}}
  & Full   &   0 & 0 &  0 & 0 \\
  & $-P_1$ & 100 & 0 &  0 & 0 \\
  & $-P_2$ &   0 & 9 &  0 & 0 \\
  & $-P_3$ &   0 & 0 & 41 & 0 \\
  & $-P_4$ &   0 & 0 &  0 & 0 \\
\bottomrule
\end{tabular}
\end{minipage}
\hfill
\begin{minipage}[t]{0.49\textwidth}
\centering
\refstepcounter{table}
\label{tab:sensitivity}
\textbf{Table~\thetable.} Extractor sensitivity: ASR (\%) per family under category
drop-out, and false-positive rate (FPR, \%) on benign traffic under over-typing, at noise rate $p$.
\vspace{0.4em}

\setlength{\tabcolsep}{3pt}
\begin{tabular}{llrrrrr}
\toprule
Case study & $p$ & $A_1$ & $A_2$ & $A_3$ & $A_4$ & FPR \\
\midrule
\multirow{4}{*}{\textsc{Shopper}}
  & $0\%$  &  0 & 0 &  0 &  0 &  0 \\
  & $10\%$ &  5 & 0 &  3 &  2 & 16 \\
  & $25\%$ & 12 & 0 & 10 & 18 & 24 \\
  & $50\%$ & 38 & 0 & 28 & 20 & 54 \\
\midrule
\multirow{4}{*}{\textsc{Airline}}
  & $0\%$  &  0 & 0 &  0 &  0 &  0 \\
  & $10\%$ &  7 & 0 &  2 &  3 &  8 \\
  & $25\%$ & 22 & 0 &  8 & 12 & 20 \\
  & $50\%$ & 37 & 0 & 20 & 16 & 36 \\
\midrule
\multirow{4}{*}{\textsc{MedAgent}}
  & $0\%$  &  0 & 0 &  0 &  0 &  0 \\
  & $10\%$ & 10 & 0 &  2 & 12 &  2 \\
  & $25\%$ & 28 & 0 & 13 & 20 & 16 \\
  & $50\%$ & 42 & 0 & 34 & 48 & 34 \\
\midrule
\multirow{4}{*}{\textsc{FinAssist}}
  & $0\%$  &  0 & 0 &  0 &  0 &  0 \\
  & $10\%$ & 10 & 0 &  2 &  0 &  0 \\
  & $25\%$ & 25 & 0 & 24 & 33 &  0 \\
  & $50\%$ & 53 & 0 & 45 & 60 &  0 \\
\bottomrule
\end{tabular}
\end{minipage}

\end{table*}

%% can we have more insights to add 
\vspace{.2cm}
\noindent\textbf{Discussion}
\label{sec:eval-synthesis}

\noindent The evaluation suggests that session history is central to compliance checking for tool-using agents. In the unmonitored setting, the agent produced violations across consent, purpose limitation, data minimization, and erasure scenarios. Stateless baselines were able to catch some violations involving obvious sensitive fields, but they missed cases where the relevant condition depended on earlier dialogue events, such as whether consent had been granted or whether an erasure request had already occurred.

The ablation results also show that the four predicates are not interchangeable. Each one guards a different GDPR principle, although some overlap appears between consent and purpose-limitation cases. This overlap is useful in practice, but it also shows that the policy design must be explicit about which legal basis and purpose apply to each tool call. Overall, the results support the use of trace-based monitoring as a practical enforcement layer for AI agents, while also showing that its reliability depends on accurate observation of natural-language and tool-call events. %natural-language and tool-call events.

In practice, \sys runs as an in-process interceptor. It does not require changes to the model or tool implementations, blocks non-compliant tool calls before execution, and writes an append-only JSONL audit log for review. A new workflow can be added by editing the policy manifest, tool-to-purpose bindings, and extractor patterns, rather than changing the monitor code.

The same trace-based structure can support additional rules, but the results also show where the approach is fragile. The monitor depends on the quality of the event annotations, especially the category extractor. The sensitivity results show that missed categories can lead to missed violations, while over-assigned categories can increase false positives. A production deployment should therefore validate the extractor on domain-specific traffic and consider fail-closed behavior for high-risk data categories. Broader scalability should also be tested under many concurrent sessions, larger manifests, and longer traces.

\vspace{.2cm}
\noindent\textbf{Threats to Validity}
\label{sec:threats}

\noindent \emph{Generalization.} The four case studies span three lawful bases and four domains, and the ordering (monitor $\gg$ regex $\gg$ none) and the headline $0\%$ ASR / $0\%$ FPR are stable across them. Absolute ASR, realization, and FPR numbers are specific to \textsf{gpt-4o-mini} at $n{=}50$ on one backend and should not be read as estimates for other models or workflows; a weaker- or stronger-aligned model would shift them, and the ablation and sensitivity magnitudes with them, while leaving the monitor's $0\%$ ASR/FPR unchanged.

\noindent\emph{Interpretation and extraction.} Our predicates capture one reasonable reading of GDPR Arts.~5--7 and 17, not the only one. They are configurable, so a deployment that reads a principle differently can replace them without changing the monitor. A second limitation is the extractor, we use a regex stand-in for a production PII classifier. RQ3 measures how the monitor degrades as this extractor drops categories, but it does not reproduce every way a real classifier fails, such as over-typing, language drift, or span errors. We keep four such cases as expected-failure tests in CI so the gap stays visible.

%%think baout more future work

%==================================================================
\section{ Conclusion}
\label{sec:discussion}
This work proposed a runtime verification framework for GDPR compliance in AI agents, combining formal predicates with an
attack driver, and evaluated it on four case studies covering retail support, airline customer service, clinical-data access,
and KYC/AML compliance (the last under Art.~6(1)(c) legal obligation
rather than consent). Under exact extraction the monitor drives attack-success and false-positive rates to zero on every family in both the generated and real-corpus registers; under 10\% per-category extractor noise the attack-success rate stays at most 12\%, below the baselines we compare against, while false positives rise to at most 16\%. The
per-workflow JSONL audit log the harness emits in audit mode is the
file a compliance reviewer would ask for. As future work, we plan to
strengthen the category extractor and validate it on domain-specific traffic, since our results indicate the monitor's reliability depends
more on extraction quality than on the policy logic.
%==================================================================

\begin{comment}
    and KYC/AML compliance (the last under Art.~6(1)(c) legal obligation
rather than consent). Under exact extraction the monitor drives attack-success and false-positive rates to zero on every family in both the generated and real-corpus registers; under 10\% per-category extractor noise the attack-success rate stays at most 12\%, below the baselines we compare against, while false positives rise to at most 16\%. The
per-workflow JSONL audit log the harness emits in audit mode is the
file a compliance reviewer would ask for. As future work, we plan to
strengthen the category extractor and validate it on domain-specific traffic, since our results indicate the monitor's reliability depends
more on extraction quality than on the policy logic.
\end{comment}
%%%%%%%%
{\footnotesize
\bibliographystyle{splncs04}
\bibliography{references}
}

\end{document}